\documentclass[fleqn,usenatbib]{mnras}

\usepackage{newtxtext,newtxmath}

\usepackage[T1]{fontenc}

\DeclareRobustCommand{\VAN}[3]{#2}
\let\VANthebibliography\thebibliography
\def\thebibliography{\DeclareRobustCommand{\VAN}[3]{##3}\VANthebibliography}

\usepackage{graphicx}	

\usepackage{subcaption}
\usepackage{float}
\usepackage{graphicx}
\usepackage{caption}
\usepackage{lipsum}
\usepackage{makecell}
\usepackage{orcidlink}
\usepackage[export]{adjustbox}

\usepackage{academicons}
\definecolor{orcidlogocol}{HTML}{A6CE39}
\newcommand{\sigmaRM}{$\sigma_{\rm RM}\,$}
\newcommand{\sigmapRM}{$\sigma^{\prime}_{\rm RM}$}
\newcommand{\RMunits}{rad$\,$m$^{-2}$}

\def\code#1{\texttt{#1}}

\title[Depolarisation of one-off ASKAP FRBs]{Searching for the spectral depolarisation of ASKAP one-off FRB sources}

\author[Uttarkar et al.]{
Pavan~A.~Uttarkar$^{1}$\orcidlink{0000-0002-2346-6853}\thanks{E-mail: puttarkar@swin.edu.au}, 
R.~M.~Shannon$^{1}$\orcidlink{0000-0002-7285-6348},
K.~Gourdji$^{1}$\orcidlink{0000-0002-0152-1129},
A.~T.~Deller$^{1}$\orcidlink{0000-0001-9434-3837},
C.~K.~Day$^{2}$\orcidlink{0000-0002-8101-3027},
S. Bhandari$^{3,4,5}$\orcidlink{0000-0003-3460-506X}
\\
$^{1}$Centre for Astrophysics and Supercomputing, Swinburne University of Technology, Hawthorn, VIC\\
$^{2}$Department of Physics, McGill University, Montreal, Quebec H3A 2T8, Canada\\
$^{3}$ASTRON, Netherlands Institute for Radio Astronomy, Oude Hoogeveensedijk 4, 7991 PD Dwingeloo, The Netherlands\\
$^{4}$Joint institute for VLBI ERIC, Oude Hoogeveensedijk 4, 7991 PD Dwingeloo, The Netherlands\\
$^{5}$Anton Pannekoek Institute for Astronomy, University of Amsterdam, Science Park 904, 1098 XH, Amsterdam, The Netherlands
}

\date{Accepted XXX. Received YYY; in original form ZZZ}

\pubyear{2015}

\begin{document}
\label{firstpage}
\pagerange{\pageref{firstpage}--\pageref{lastpage}}
\maketitle

\begin{abstract}
Fast Radio Bursts (FRBs) are extragalactic transients of (sub-)millisecond duration that show wide-ranging spectral, temporal, and polarimetric properties. The polarimetric analysis of FRBs can be used to probe intervening media,  study the emission mechanism, and test possible progenitor models. In particular, low frequency depolarisation of FRBs can identify dense, turbulent, magnetised, ionised plasma thought to be near the FRB progenitor. An ensemble of repeating FRBs has shown low-frequency depolarisation. The depolarisation is quantified by the parameter \sigmaRM, which correlates with proxies for both the turbulence and mean magnetic field strength of the putative plasma. However, while many non-repeating FRBs show comparable scattering (and hence inferred turbulence) to repeating FRBs, it is unclear whether their surrounding environments are comparable to those of repeating FRBs. To test this, we analyse the spectro-polarimetric properties of five one-off FRBs and one repeating FRB, detected and localised by the Australian Square Kilometer Array Pathfinder. We search for evidence of depolarisation due to \sigmaRM and consider models where the depolarisation is intrinsic to the source. We find no evidence (for or against) the sample showing spectral depolarisation. Under the assumption that FRBs have multipath propagation-induced depolarisation, the correlation between our constraint on \sigmaRM and RM is consistent with repeating FRBs only if the values of \sigmaRM are much smaller than our upper limits. The observations provide further evidence for differences in the environments and sources of one-off and repeating FRBs.

\end{abstract}




\section{Introduction}

Fast Radio Bursts (FRBs) are (sub-)millisecond duration transients of extragalactic origin. The first FRB was  discovered in archival Parkes data \citep{Lorimer}. Since the first reported burst, there have been more than 600 FRB detections published  \citep[e.g.,][]{CHIME_Catalogue_1}.

Of those, only $\sim$50 FRBs are currently known to repeat, with FRB 20121102A and FRB 20180916B showing periodic activity \citep{FRB180916B_periodic, Rajwade}. Whether all FRBs can be eventually observed to repeat  is  unclear \citep{Caleb_2019}.
FRBs have been observed over a wide range of frequencies, in bands from $110-180$\,MHz \citep{180916B} to $4-8$\,GHz \citep{Gajjar}.
While FRBs were initially speculated to be associated with cataclysmic events \citep{Thronton}, this thinking had to be updated after the discovery of the first repeating FRB \citep{20121102A_repeater}. An early analysis of four bursts detected from the High Time Resolution Universe (HTRU)  Survey using the Parkes 64-m radio telescope suggested the FRB volumetric occurrence rate above a fluence of $3$ Jy\,ms to be $\sim$$10^4$ sky$^{-1}\,$day$^{-1}$ \citep{Thronton}. An analysis by \cite{Ravi_2019} considering the CHIME/FRB detections shows the FRB volumetric occurrence rate to be higher than the event rate for cataclysmic source classes such as supernova explosions or neutron star mergers, concluding that a significant fraction of FRBs could be repeating sources.

In addition to representing a new astrophysical phenomenon of unknown origin, FRBs promise to be useful as cosmological probes. The interferometric localisations of one-off FRB sources to their host galaxies \citep[e.g.,][]{Bannister_2019,Prochaska,Cho,Day_2020} have enabled the measurement of the cosmic baryon density in the low-redshift Universe \citep{Macquart}. Additionally, localised FRB sources from the Australian Square Kilometer Array Pathfinder (ASKAP), along with Parkes-detected FRBs, have provided an independent measurement of Hubble's constant \citep{Hubble}. In the future, a more precise measurement with an uncertainty of  $\pm 2.5$\,km\,s$^{-1}$\,Mpc$^{-1}$ may be possible with a larger sample of localised FRBs \citep{Hubble}.

Repeating and (apparently) non-repeating FRBs can show different characteristic spectral and temporal properties. This includes linear frequency drift of intra-burst emission with time ("sad-trombone" structure) observed in repeating FRBs \citep[e.g.,][]{Hessels,R2_drift}. There is also evidence that the bursts from repeating FRBs are generally of longer duration than non-repeating FRBs and that repeating FRBs show band-limited structure \citep[e.g.,][]{Gourdji19,CHIME_Catalogue_1,Pravir_2021,Pleunis21}. Some repeating FRBs also exhibit an increase in fractional bandwidth with frequency \citep{Bethapudi}.

A small fraction of repeating FRBs also exhibits drastic variations in spectro-polarimetric characteristics, such as the extremely narrow-banded burst observed from the repeating source FRB 20190711A \cite[][]{Pravir_2021}. Initial 5-GHz polarimetric studies of the first known repeating FRB 20121102A reported a high degree of linear polarisation ($\sim$100\%) along with a flat polarisation position angle (PPA) \citep{Michilli}. However, subsequent reported observations of repeaters, such as FRB 20201124A \citep{Pravir_2022} and FRB 20180301A \citep{Rui_Luo} indicate more diverse polarimetric behaviour, with the PPA varying across the burst envelope. In comparison, the sample of non-repeaters with polarimetric information displays a variety of polarisation fractions, with the majority showing high degrees of linear polarisation \citep[>$90$\%;~e.g.,][]{Cho, Day_2020}. Some non-repeating FRBs also show a relatively high degree of circular polarisation \citep[$\sim$$5$\%;~e.g.,][]{Bhandari, Day_2020}, while others
appear to exhibit sub-burst to sub-burst variation in the polarisation fraction along with varying Rotation Measures (RM) \citep{Cho, Day_2020}. 

Repeating and non-repeating FRBs also show differences in RM, with some repeating FRBs showing extremely large RM magnitudes, such as the first detected repeater FRB~201211202A \citep{Michilli}, which has a measured RM of $\sim$$10^5~$\RMunits. This FRB also shows a secular decline in RM ($\sim$15 \%~/~year) as well as short-term variations of up to $\sim$1 \%~/~week \citep[][]{Michilli,hilmarsson21}. FRB 20190520B displays an even larger RM variation from 10$^4$\RMunits\ to $\sim$ $-1.6\times10^4$ \RMunits\ over approximately six months suggesting a reversal in the magnetic field. Such variation in RM shown in FRBs 20190520B and  20121102A has been interpreted as arising from a complex turbulent circumburst magnetoionic environment \cite[][]{Dai_S, Anna-Thomas_20190520B}.
The RM magnitude and variations can be used to develop models for the surrounding environment. 
For instance, the detected reversal in the magnetic field reported by \cite{Dai_S} was speculated to be caused by the presence of a highly magnetised stellar or black-hole companion. 
In contrast, non-repeating FRBs have been observed to have less extreme rotation measures (less than a few hundred rad\,m$^{-2}$), consistent with passage through Milky-Way-like columns in the interstellar medium (ISM)\footnote{For obvious reasons, long term rotation measure variations have not been observed in non-repeating FRBs.}  \cite[][]{mannings2023}.

Another potential diagnostic of FRB environments arises from the spectral depolarisation of burst emission.
Early observations showed that the linear polarisation fraction of the Crab nebula was lower than expected and decreased at lower frequencies \citep{Gardner}. The variation of the linear polarisation fraction as a function of frequency was explained by \cite{Burns_1966} to be due to the random orientation of the magnetic field, leading to Faraday rotation in its shell. The strength of the effect is quantified by the scatter RM measure, \sigmaRM. Similar depolarisation behaviour was recently reported by \cite{Feng} in repeating FRBs.  Using observations from several different telescopes across a frequency range of 115 MHz to 4600 MHz, their sample also shows a clear trend of decreasing linear polarisation fraction with decreasing frequency, with reported \sigmaRM from $0.12$ \RMunits\ to 218 \RMunits. 
The depolarisation reported was quantified using multipath propagation due to the foreground scattering screen, which includes the host galaxy and Milky Way. This is different from the depolarisation which has been seen in extended sources, which can be explained due to the differential RM in the source itself \citep{Burns_1966}.
A correlation was also reported between the \sigmaRM, RM and scatter-broadening time $\tau_s$ \citep{Feng}. 
This inferred depolarisation behavior provides a diagnostic of intervening media and can be used to constrain possible progenitor and emission mechanism models.
Notably, the two sources with the highest values of \sigmaRM, FRBs 20121102A \citep[30.9$\pm$0.4 \RMunits;~][]{Chatterjee, Feng} and 20190520B \citep[218 $\pm$ 10.2 \RMunits;~][]{Nui, Feng}, are associated with persistent radio sources. 

It is unclear if spectral depolarisation is present in non-repeating FRBs. Measuring this effect is challenging in non-repeating FRBs as the
search data streams are constrained to modest observing bandwidths, and low time and frequency resolution.
These streams might also be affected by intra-channel depolarisation and dispersion smearing.
Some search systems include voltage buffers that  provide higher time and spectral resolution, but these are usually still bandwidth limited.
An investigation of the effects of depolarisation among non-repeating FRBs would provide insight into the dichotomy of repeating and non-repeating FRBs and potentially serve as a diagnostic of the properties of their respective circumburst media.

The Commensal Real-Time ASKAP Fast Transient (CRAFT) survey system currently utilises two parallel data streams: a low time resolution stream in Stokes I to search for FRBs and a $\sim$3.1-s voltage buffer, which is triggered if an FRB candidate is detected.
These voltages, saved for individual antennas, are then used for interferometric localisation and to produce high-time resolution data products \citep[e.g.,][]{Bannister_2019, Cho, Day_2020, Macquart}. In this paper, we study ASKAP reported FRBs, among which FRB 20180924B \citep{Bannister_2019}, FRB 20190611B, FRB 20190102C, FRB 20190608B and FRB 20191001A are one-off FRBs \citep{Bhandari, Day_2020, Macquart}.  We also examine the repeating source FRB 20190711A \citep{Day_2020}. We search for depolarisation in individual sub-bursts as well as the integrated burst for all FRBs. We also compare three different depolarisation models using the set of six FRBs to investigate
apparent correlations between  \sigmaRM, RM and $\tau_s$. We describe the data selection and  models for depolarisation in Section \ref{sec:Methodology} and the results in Section \ref{sec:Results}. We conclude with a discussion of the potential models and implications for the populations of repeaters and non-repeaters in Section \ref{sec:Discussion}.

\section{Methodology}
\label{sec:Methodology}
\subsection{Data selection and description}
\label{sec:data_description}
The polarimetric analysis is performed on a set of five localised ASKAP one-off FRB sources and one localised repeating ASKAP FRB source. These FRB datasets were detected by the CRAFT real-time detection system and were first reported in \cite{Bannister_2019}, \cite{Bhandari}, and \cite{Macquart}. After the bursts were detected, dual-polarisation voltage buffers were downloaded from individual antennas. These voltage buffers were then calibrated for flux, phase, and polarisation and then coherently analysed to study burst spectro-temporal polarimetry \citep{Day_2020}.
For FRBs 20180924B, 20190608B,  20190102C,  20190611B, and 20190711A, we used image domain, de-dispersed data sets described in \cite{Day_2020}. For FRB 20191001A, we used the High Time Resolution (HTR) de-dispersed complex voltages presented in \cite{Bhandari}, using methods described in \cite{Cho}.

The data used to analyse FRBs 20180924B, 20190102C,  20190608B,  20190611B, and  20190711A have a recorded bandwidth of 336 MHz, with $\sim$4 MHz of channel resolution each, at a center frequency of 1272 MHz. The data used to analyse FRBs 20180924B and  20191001A have  central frequencies  of 1320 MHz and 824 MHz, respectively. For FRB 20191001A, there was only 144 MHz of recorded bandwidth because of the latency in the detection pipeline and short ($3.1$\,s) duration of the voltage buffer, which caused the FRB to be overwritten in the high frequency part of the band.

\begin{figure*}
     \centering
     \begin{subfigure}[b]{0.3\textwidth}
         \centering
         \includegraphics[width=\textwidth]{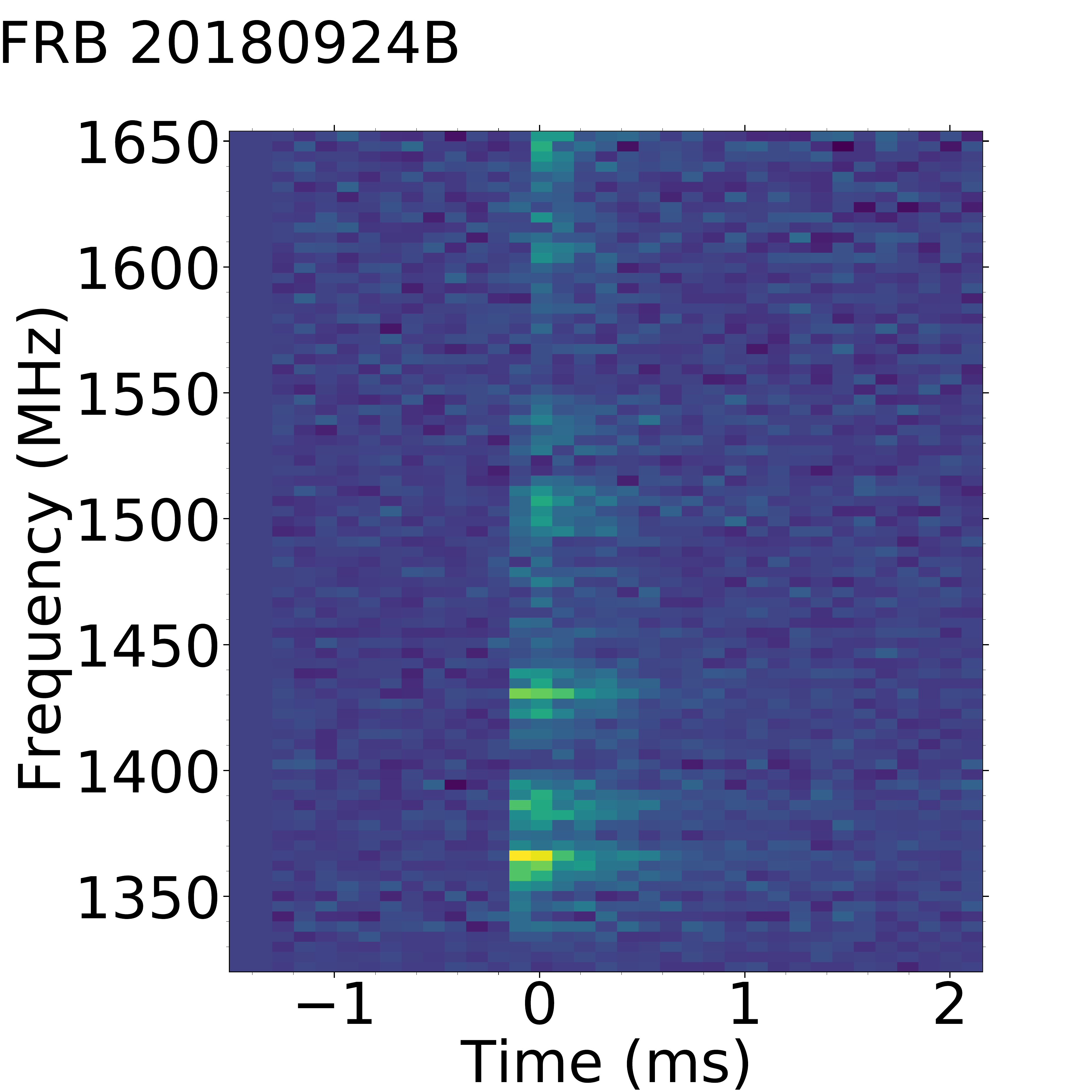}
         \caption{}
         \label{fig:180924_spec}
     \end{subfigure}
     \begin{subfigure}[b]{0.3\textwidth}
         \centering
         \includegraphics[width=\textwidth]{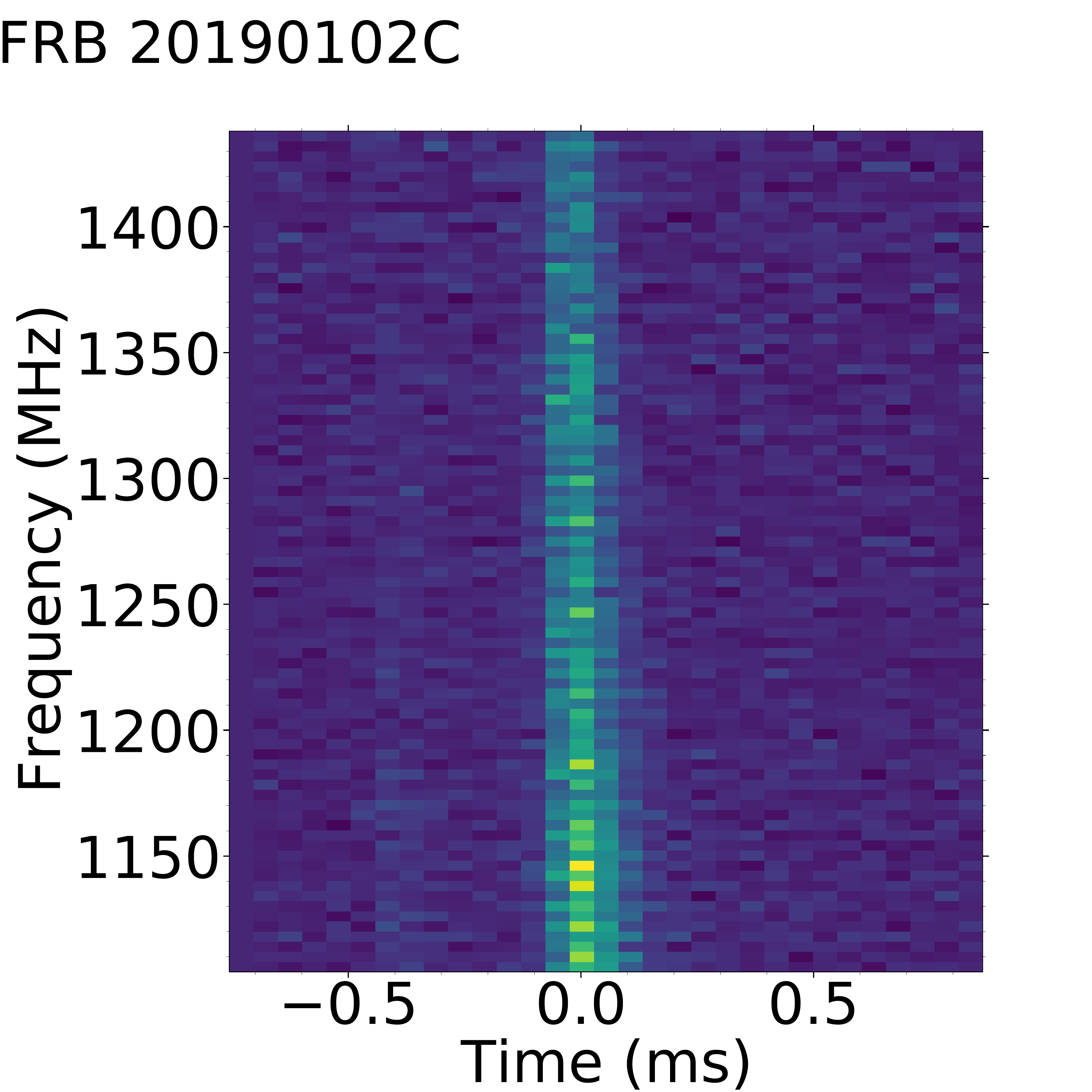}
         \caption{}
         \label{fig:190102_spec}
     \end{subfigure}
     \begin{subfigure}[b]{0.3\textwidth}
         \centering
         \includegraphics[width=\textwidth]{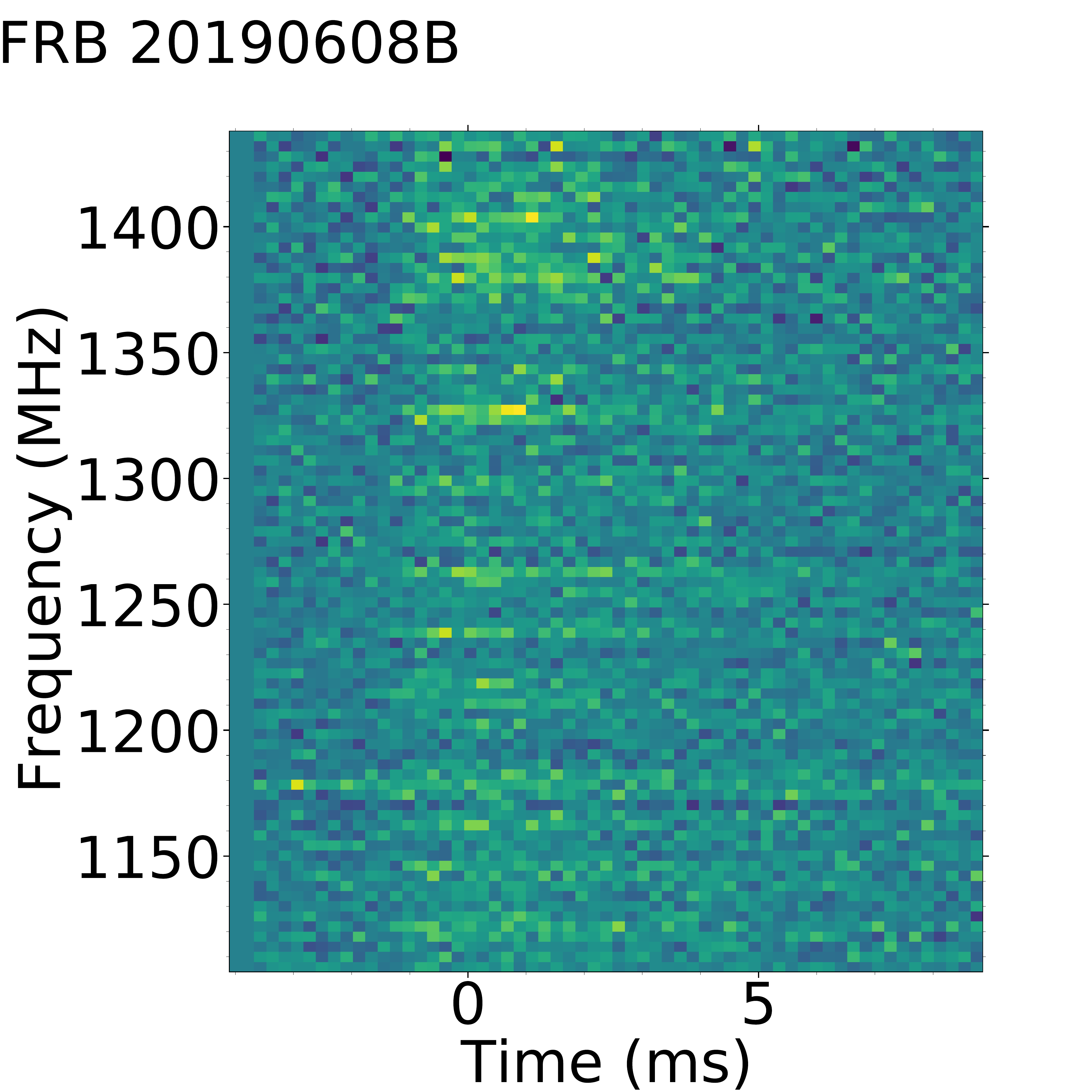}
         \caption{}
         \label{fig:190608_spec}
     \end{subfigure}\\

     \centering
     \begin{subfigure}[b]{0.3\textwidth}
         \centering
         \includegraphics[width=\textwidth]{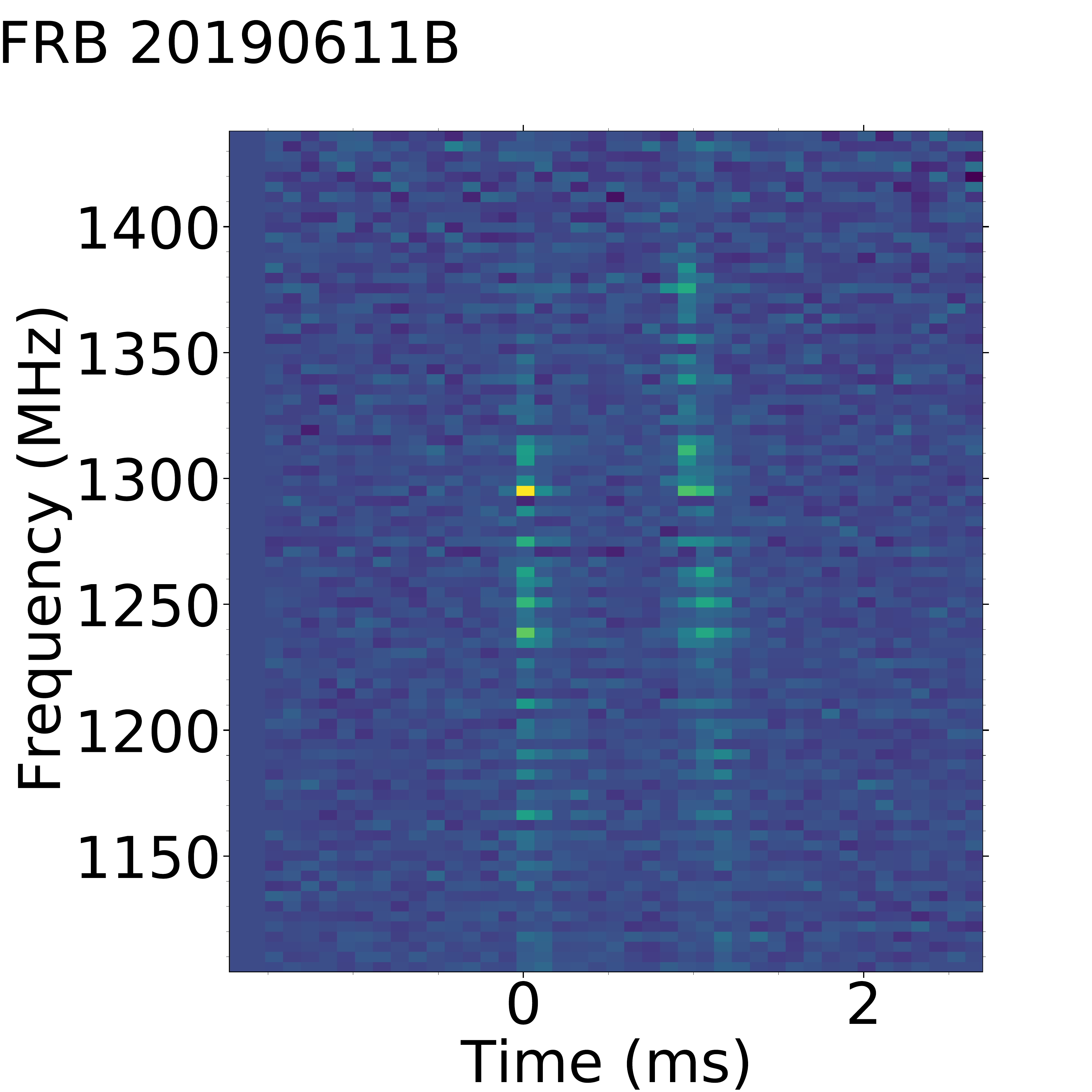}
         \caption{}
         \label{fig:190611_spec}
     \end{subfigure}
     \begin{subfigure}[b]{0.3\textwidth}
         \centering
         \includegraphics[width=\textwidth]{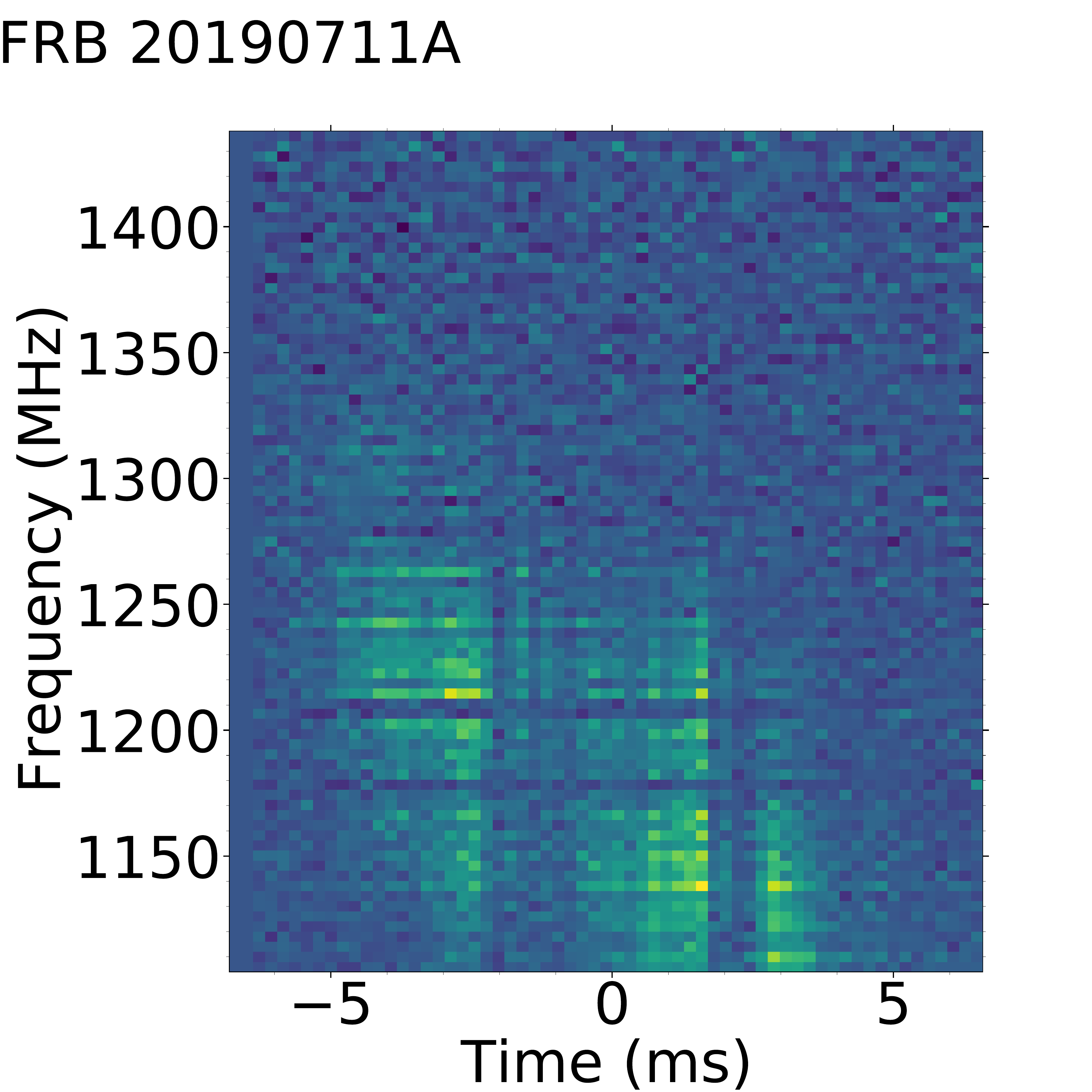}
         \caption{}
         \label{fig:190711_spec}
     \end{subfigure}
     \begin{subfigure}[b]{0.3\textwidth}
         \centering
         \includegraphics[width=\textwidth]{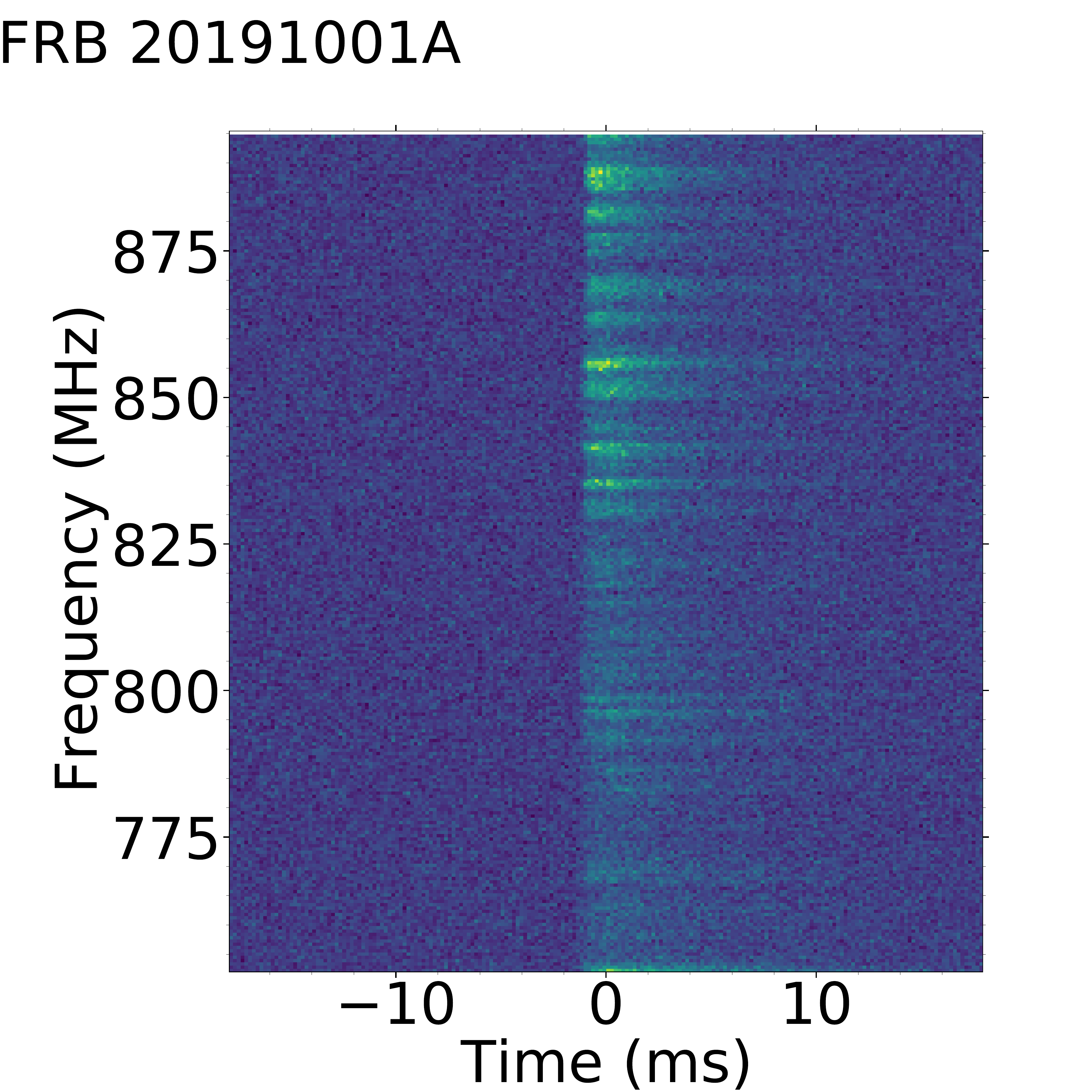}
         \caption{}
         \label{fig:191001_spec}
     \end{subfigure}
        \caption{Stokes-I dynamic spectra of the bursts used for the depolarisation analysis. FRBs 20180924B,  20190102C,  20190608B,  20190611B, and  20190711A are shown with a spectral resolution of ~4 MHz and temporal resolution of 0.108 ms, 0.054 ms, 0.261 ms, 0.108 ms, and 0.216 ms respectively. FRB 20191001A has a spectral resolution of ~0.5 MHz and a temporal resolution of 0.18 ms.}
        \label{fig:Stokes_I_dynamic}
\end{figure*}

The dynamic Stokes-I spectra of the bursts are shown in Figure \ref{fig:Stokes_I_dynamic}.  
Among our sample, FRBs 20190611B,  20190711A, and  20190102C show distinct sub-bursts. FRB 20190102C has one faint sub-burst before the main pulse that shows a significantly different polarisation fraction relative to the main burst but with a similar RM \cite[][]{Day_2020}. The repeating source FRB 20190711A shows three characteristic sub-bursts, all having similar RMs and showing complex pulse morphology \cite[][]{Day_2020}. 
With the exception of FRB 20191001A, all FRBs have a high fractional linear polarisation \citep[>$90$\%;][]{Bhandari, Day_2020}.

\subsection{Measuring linear polarisation fraction}
\label{sec:measure_L_I}
Faraday rotation is the frequency dependent rotation of plane of linear polarisation induced by the magnetic field component parallel to the line of sight of the observer to the intervening cold magnetised plasma. The magnitude of the effect is quantified by the rotation measure (RM). The change in the polarisation position angle (PPA) is given by
\begin{equation}
    {\rm PPA} = {\rm RM_{obs}} (\lambda^2 - \lambda_o^2),
	\label{eq:PPA}
\end{equation}
where $\lambda$ is the wavelength, $\lambda_o$ is the wavelength of the center frequency of the observing band and RM$_{\rm obs}$ is defined to be
\begin{equation}
    {\rm RM}_{\rm obs} = \frac{e^3}{2\pi m_e^2 c^4}\int_{d}^{0}\frac{n_e B_{||}}{(1+z)^2} dl,
	\label{eq:RM_calulation}
\end{equation}
where $d$ is the distance to the source of emission, $m_e$ is the electron mass, $e$ is the charge of the electron, $B_{||}$ is the magnetic field parallel to the line of sight and $z$ is the redshift of the plasma. The individual Dispersion Measure (DM) and RM of the bursts reported in \cite{Bhandari} and \cite{Day_2020} are listed in Table \ref{tab:Polarization_parameters}.
We calculate the fraction of linear polarisation of individual sub-bursts and the integrated burst (integration across all the sub-bursts) from the polarisation calibrated data set for our sample. (For a rigorous description of the analysis, refer \cite{Bhandari} and \cite{Day_2020}.)

\begin{table}
\begin{tabular}{ccrr} 
	\hline
	 \rule{0pt}{2ex}  TNS Name & Sub-burst &\thead{DM \\(pc cc$^{-3}$)}&\thead{Rotation Measure\\ (\RMunits)}  \\
	\hline
	FRB 20180924B  &  & 362.2     &  22$\pm$2    \\   
	\\
	FRB 20190102C  &  & 364.538   & -105$\pm$1   \\
	\rule{4pt}{0ex}&  
	    sub-burst 0 &           & -128$\pm$7   \\
    \rule{4pt}{0ex}&
	    sub-burst 1 &           & -105$\pm$1   \\
	\\
	FRB 20190608B  &  & 339.79    & 353$\pm$2    \\
	\\
	FRB 20190611B  &  & 332.60    & 20$\pm$4     \\
	\rule{4pt}{0ex}&
	    sub-burst 0 &           & 19$\pm$4     \\
	\rule{4pt}{0ex}&
	    sub-burst 1 &           & 12$\pm$2     \\
	\\
	FRB 20190711A  &  & 587.87  & 9$\pm$2      \\
	\rule{4pt}{0ex}&
	    sub-burst 0 &           & 10$\pm$4     \\
	\rule{4pt}{0ex}&
	    sub-burst 1 &           & 9$\pm$3      \\
	\rule{4pt}{0ex}&
	    sub-burst 2 &           & 12$\pm$6     \\
	\\
	FRB 20191001A  &  & 506.92   & 55.5$\pm$9   \\
	\hline
\end{tabular}
\caption{Rotation Measure and Dispersion Measure for FRBs in our sample. For bursts with subcomponents we list both the quantities derived from those and the entire burst.}
\label{tab:Polarization_parameters}
\end{table}

To measure the linear polarisation fraction of the bursts, we first de-rotate the calibrated spectra to account for burst RM: 
\begin{equation}\label{eq:L_correct}
\begin{split}
    U_{\rm cor} &= - Q_{\rm cal}\, {\rm cos(2\phi)} + U_{\rm cal}\,\rm sin(2\phi),\\
    Q_{\rm cor} &= Q_{\rm cal}\, {\rm cos(2\phi)} + U_{\rm cal}\, \rm sin(2\phi),\\
\end{split}
\end{equation}
where $Q_{\rm cal}$ and $U_{\rm cal}$ are the polarisation-calibrated spectra of individual bursts
and $\phi$ is the PPA.
The de-rotation is conducted using the RMs from Table \ref{tab:Polarization_parameters} for individual sub-bursts in the set of FRBs. 
After this, further averaging in frequency is performed to increase the signal-to-noise ratio (S/N) in individual spectral measurements.
We average each FRB and sub-burst differently to account for their varying S/N, as described in Section \ref{sec:Bayesian_Modeling}. The total linear polarisation estimator $L_{\rm cal}=\sqrt{Q_{\rm cor}^2+U_{\rm cor}^2}$ is a biased estimate due to the presence of noise, especially at low to intermediate S/N  \citep{Everett_Wiseberg}. To obtain an unbiased estimate of the polarisation fraction, de-biasing is performed, following Equation \ref{eq:debais}, originally presented in \cite{Wardle_Kronberg}:
\begin{equation}
    L_{\rm de-bias}  =  
    \begin{cases}
        \sigma_{I}\,\sqrt{\left(\frac{L_{\rm cal}}{\sigma_{I}}\right)^2 - 1}\,,& \text{if } \frac{L_{\rm cal}}{\sigma_I}\geq 1.57\beta \\
    0,              & \text{otherwise},
    \end{cases}
	\label{eq:debais}
\end{equation}
where $L_{\text{cal}}$ is the Linear polarisation calculated from the calibrated Stokes-Q and Stokes-U spectra, and $\sigma_I$ is the uncertainty in the Stokes I spectrum. We follow the de-biasing method used in \cite{Everett_Wiseberg} and choose a higher threshold of $\beta$ = 2 for our analysis, to account for low S/N measurements, due to the band-limited structure in some of the bursts.

\subsection{Bayesian modelling for match-filtering}
\label{sec:Bayesian_Modeling}
To maximise the S/N in an individual Stokes spectrum, we use a match-filtering approach to average over the burst envelope. On average, the match filtering approach provides an improvement of $\sim$49\% in the polarised SNR, relative to a simple boxcar (i.e., non-weighted average). The improvement is more pronounced for FRBs 20180924B, 20190102C, 20190608B, 20190611B, and 20190711A, which have HTR image domain datasets - compared to FRB 20191001A, which has HTR beamformed data available. Using a Bayesian framework, we model the burst envelopes assuming either a  Gaussian or a Gaussian convolved with an exponential if the burst shows evidence for scatter broadening. 

In the case of a Gaussian pulse, we model the burst to be
\begin{equation}\label{eq:model_gauss}
\begin{split}
    W_{\rm pulse}(t)        = A\,\exp \left({-\frac{(t-t_0)^2}{2\sigma^2}} \right)\\
\end{split},
\end{equation}
where $A$ is the amplitude of the pulse,  $t_0$ is the centre of the pulse and the pulse width (the full width at half maximum) is $2\sigma\sqrt{2 \ln 2}$.

In the case of a scatter-broadened pulse, we assume the pulse to be a Gaussian convolved with an exponential, pulse-broadening function (PBF),
\begin{equation}\label{eq:model_expgauss}
\begin{split}
    {\rm PBF}(t) = 
        \begin{cases}
        \exp \left( {-\frac{t}{\tau}} \right) & \text{if } t>0\\
    0,              & \text{otherwise},
    \end{cases}
\end{split}
\end{equation}
where $\tau$ is the scattering time. The weights are calculated for Gaussian and a Gaussian convolved exponential functions as defined above.
A weighted average is performed with the weights estimated for individual time bins of the de-rotated Stokes spectrum at coarser frequency resolution of 33.6 MHz, 28 MHz, 67.2 MHz, 33.6 MHz, 56 MHz, and 14.4 MHz for FRBs 180924B, 20190102C, 20190608B, 20190611B, 20190711A, and 20191001A respectively, to increase the S/N. 

We use exponentially convolved Gaussian models for FRBs 20180924B,  20190608B, and  20191001A. For FRBs 20190102C and  20190611B, Gaussian and exponentially convolved Gaussian models are used for different components (we refer the readers to \cite{Day_2020}, \cite{Bhandari}, \cite{Qiu_2020} for a detailed description of the pulse-broadening measurements). To fit these models, we use a Gaussian likelihood (and assume the noise in our data is Gaussian distributed):
\begin{equation}
\begin{split}
     \textit{L}(P_i| P_M , \sigma) =  \prod_i^{N_f} \frac{1}{\sqrt{2\pi\sigma^2}}\,\,\exp\left(-\frac{(P_i-P_{M,i})^2}{2\sigma^2}\right),
    \label{eq:posterior_likelihood}
\end{split}
\end{equation}
where $P_i$ is the measured fractional polarisation data in frequency channel $i$, $P_M$ is the modelled fractional linear polarisation fraction, and $\sigma$ is the standard deviation.
We describe the models for linear polarisation fraction $P_M$ below. 
Parameter estimation is performed using the software package \code{bilby} \citep{Ashton} using the \code{DYNESTY} \citep{DYNESTY} nested sampler.
 We assume a uniform prior for $\tau$ and  fixed $t_0$. For Gaussian pulse we use the (flux-weighted) centroid of emission to define $t_0$. For a scattered pulse we define $t_0$ to be at leading edge at 10\% of the peak.

\subsection{Maximising S/N using Match Filtering}\label{sec:SNR_match_filtering}
The weights calculated for individual time bins are used to average over  frequency, to generate coarser frequency channels as given by
\begin{equation}
    P_{I, Q, U}(t, \nu) =  \sum^{\nu_1}_{\nu=\nu_0}\sum^{T}_{t=0} S_{I, Q, U}(t, \nu) \, W_{\rm pulse}(t),
    \label{eq:weighing}
\end{equation}
where $P_{I, Q, U}(t, \nu)$ are the on-pulse weighted average across  frequency, $S_{I,Q,U}(t, \nu)$ are the de-rotated Stokes I, Q, and U spectra and $W_{\rm pulse}(t)$ are the weights assigned to the individual sub-bursts for their respective time integrations. This is computed for all of the Stokes components, for both individual and integrated sub-bursts. The calculated linear polarisation fractions from the coarser frequency channels are used in our modelling. We use a top-hat average for individual sub-bursts in the case of FRB 20190711A, due to its complex pulse morphology.
The fractional linear polarisation is calculated as $L_{\rm de-bias}/P_I$, and the uncertainties in the $L_{\rm de-bias}$ are estimated using the uncertainties obtained in the Stokes I, Q, and U spectra ($\sigma_I$, $\sigma_Q$, and $\sigma_U$, respectively) determined by 
\begin{equation}
    \sigma_{L}(\nu) =  \left[ \left(\frac{P_U(\nu)}{\sqrt{P_Q^2+P_U^2}} \sigma_U\right)^2 +  \left( \frac{P_Q(\nu)}{\sqrt{P_Q^2+P_U^2}} \sigma_Q \right)^2 \right]^{\frac{1}{2}},
    \label{eq:L_error}
\end{equation}
where $P_{Q, U}$ are the weighted coarser sub-burst time-frequency averaged spectra and $\sigma_{Q, U}$ are their respective uncertainties. The uncertainties in the original Stokes spectra are derived from their respective image-domain root mean square spectra, averaged using the weights derived from the pulse profiles for individual coarse channels.
In the case of FRB\,20191001A, we use the off pulse baseline to estimate these values. 
The uncertainty in the polarisation fraction $\sigma_{L/I}$ is calculated using
\begin{equation}
    \sigma_{L/I}(\nu) =  \frac{\sqrt{\left(\sigma_L^2\,P_I^2 + L_{\rm de-bias}^2\,\sigma_I^2 \right)}}{P_I^2}. 
    \label{eq:L_I_error}
\end{equation}

The uncertainties calculated from Equation \ref{eq:L_I_error} are measured for individual sub-bursts across the averaged frequency channels. Further, likelihoods for individual models are generated using these estimates and compared individually.

\newcommand\wdt{30}

\subsection{Depolarisation: Comparison of models.}\label{sec:Depol_model}
We consider three models for the observed linear polarisation fraction of our FRB sample:
\begin{enumerate}
\item[1] Burn's law of foreground depolarisation assuming 100\% intrinsic polarisation at all frequencies;
\item[2] Burn's law assuming a constant intrinsic fractional linear polarisation at all frequencies that is <100\%;
\item[3] Constant linear polarisation fraction at all frequencies with no depolarisation.
\end{enumerate}

The linear polarisation fraction calculated for the set of FRBs is independently fitted to all three models using a Bayesian framework. The Bayesian approach allows us to compare models, through the use of Bayesian evidence.
We note that previous work only considered Case 1 \citep{Feng}.


Case 1. \textit{Burn's law of foreground depolarisation}: In this case, the depolarisation in  frequency is entirely due to the scatter in RM contributed by the line-of-sight (refer to Appendix \ref{apx:foreground_depolarisation} for discussion) and can be modelled as 
\begin{equation}
    P_{\rm Burn}(\lambda) =  \exp{\left(-2\sigma_{\rm RM}^2\lambda^4\right)},
    \label{eq:Burns_law}
\end{equation}
where \textit{$P_{\rm Burn}(\lambda)$} is the linear polarisation fraction, \sigmaRM is the scatter RM, and $\lambda$ is the wavelength of observation. 

Case 2. \textit{Modified Burn's law}: Burn's law assumes 100\% polarisation at infinite frequency. While many repeating FRBs show high degrees of linear polarisation, this need not be the case. To model lower fractional linear polarisation, we introduce  $P_{\rm int}$, which accounts for potentially lower polarisation fraction, and a modified depolarisation  parameter $\sigma^\prime_{\rm RM}$: 
\begin{equation}
    P_{\rm modBurn}(\lambda) =  P_{\rm int}\,\exp{\left(-2\sigma^{' 2}_{\rm RM}\lambda^4\right)}.
    \label{eq:Burns_law_modified}
\end{equation}

Case 3. \textit{Constant depolarisation case}:
We also consider the possibility that there is no foreground spectral depolarisation and that  depolarisation is intrinsic to the source and constant across the band.  In this case the model is
\begin{equation}
    P_{\rm no-depol}(\lambda) =  P_{\rm int},
    \label{eq:No_depolarisation}
\end{equation}
where $ P_{\rm int}$ parameterises the depolarisation.

\subsubsection{Evaluating depolarisation Models}
The marginal likelihood is calculated for the models described above for each sub-burst as well as the total integrated burst. 
These are used to determine the Bayes factor and select a preferred model.
Models 1 and 3 each have one parameter: \sigmaRM  and  $ P_{\rm int}$, respectively.
For the modified Burn's law model, there are two parameters: $ P_{\rm int}$ and  $\sigma^\prime_{\rm RM}$.  In all cases, we assume uniform priors on the  parameters.
For $\sigma^\prime_{\rm RM}$ and \sigmaRM, we assume a uniform prior between 0 and 20 \RMunits.  For $P_{\rm int}$, we assume a uniform prior between 0 and 1.

\section{Results}
\label{sec:Results}

The marginal likelihood for all the models are listed in Table \ref{tab:evidence_parameters} along with the maximum-likelihood parameters, the 95 \% CI upper limits for \sigmapRM\,, and the 1-$\sigma$ credible regions for P$_{\rm int}$ and \sigmaRM.
We calculate three Bayes factors -- $\Delta$log$_{10} E_{\rm BL}$, $\Delta$log$_{10} E_{\rm MD}$ and $\Delta$log$_{10} E_{\rm ND}$ -- to assess the pairwise preference of the models.
To assess the significance of  model preference, we follow \cite{Trotta_2008}, interpreting  log$_{10}$ evidence $\Delta E \geq$  10 to be strong evidence and $\Delta E \leq$  1 as inconclusive.  
We report constraints on \sigmaRM\ and \sigmapRM\ values for integrated bursts and sub-bursts in Table \ref{tab:evidence_parameters}. 
 We do this even when Burn's law and modified Burn's law are not favoured, as they can be used to constrain the magnetoionic properties of any foreground plasma.

\begin{figure}
        \subcaptionbox[width=1\columnwidth \label{fig:Burns_Posterior}]{}
	{\includegraphics[width=0.78\columnwidth]   {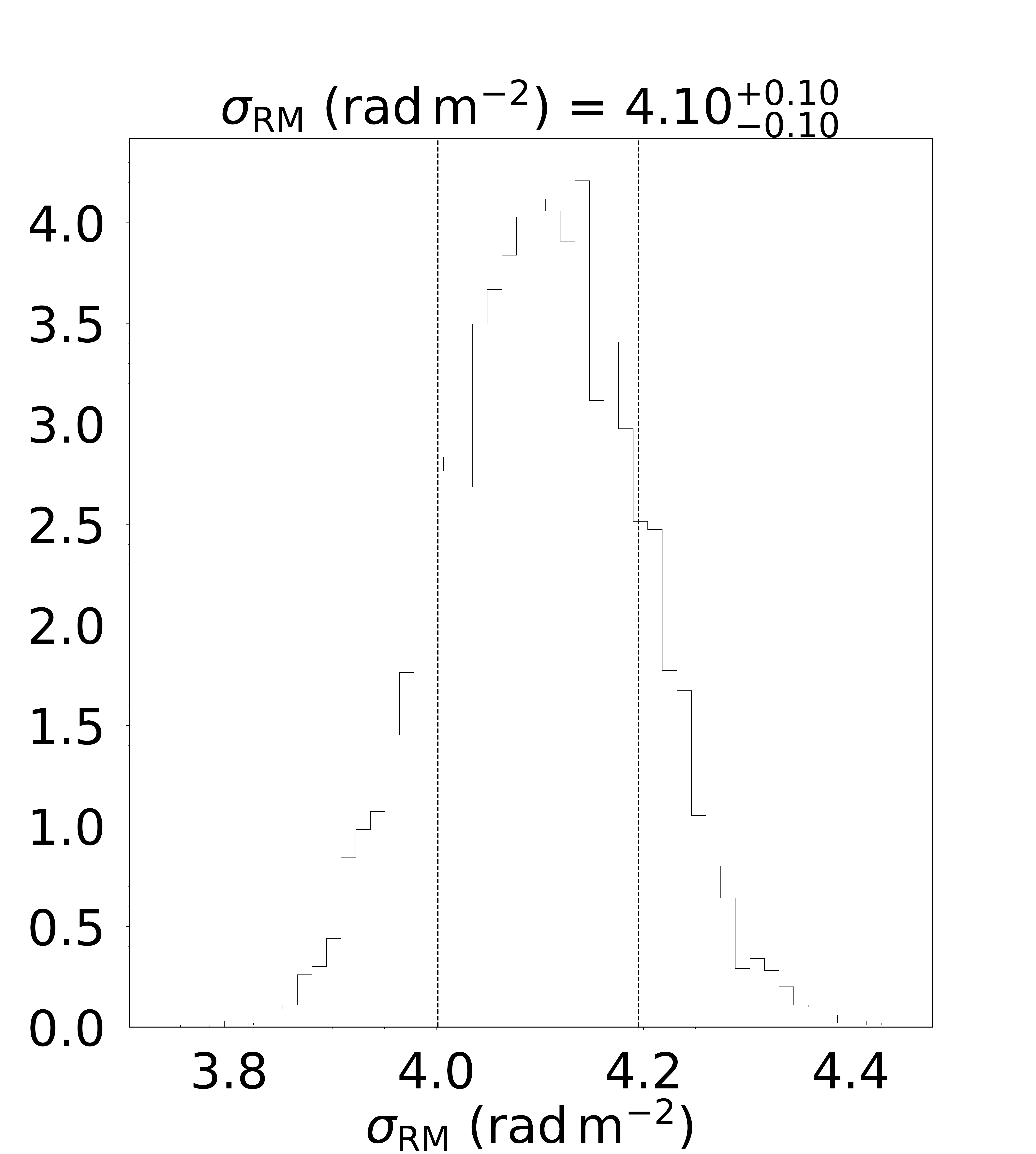}}
 
	\subcaptionbox[width=1\columnwidth \label{fig:Posterior_constPol}]{}
	{\includegraphics[width=0.78\columnwidth]{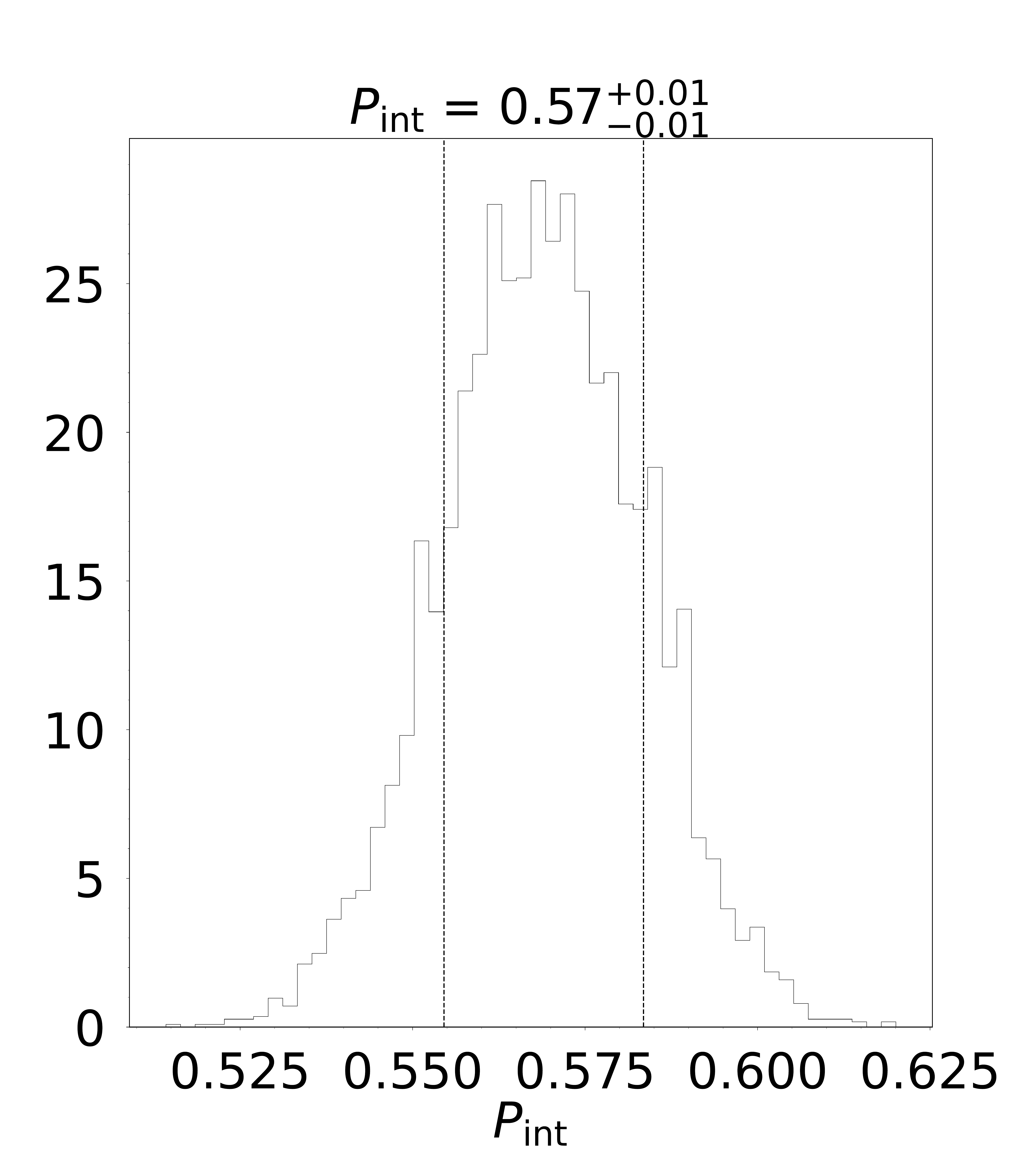}}

	\subcaptionbox[width=1\columnwidth \label{fig:low_Pol_Posterior}]{}
	{\includegraphics[width=0.78\columnwidth]{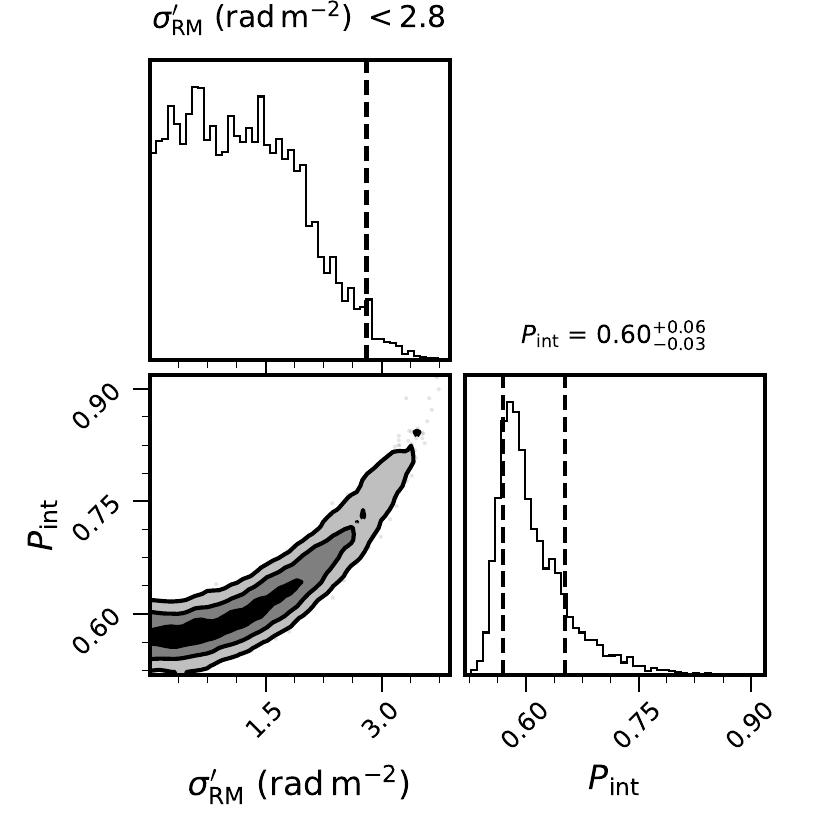}}
\end{figure}

\begin{figure}
    \ContinuedFloat  
    
    \caption{Posterior distributions for FRB 20191001A. We show the model for Burn's law of foreground depolarisation in panel (a), the constant polarisation model in panel (b) and the modified version of the Burn's law in panel (c). The dashed lines represent the 68\% uncertainty for panels (a) and (b). For panel (c), the dashed lines represent the 95\% upper limit for \sigmapRM\ and 68\% uncertainty for $P_{\rm int}$. The \sigmaRM\ derived from the unmodified Burn's law (panel (a)) is disfavoured relative to the other two models shown in panels (b) and (c).}
    \label{fig:191001_posterior}
\end{figure}

\begin{figure*}
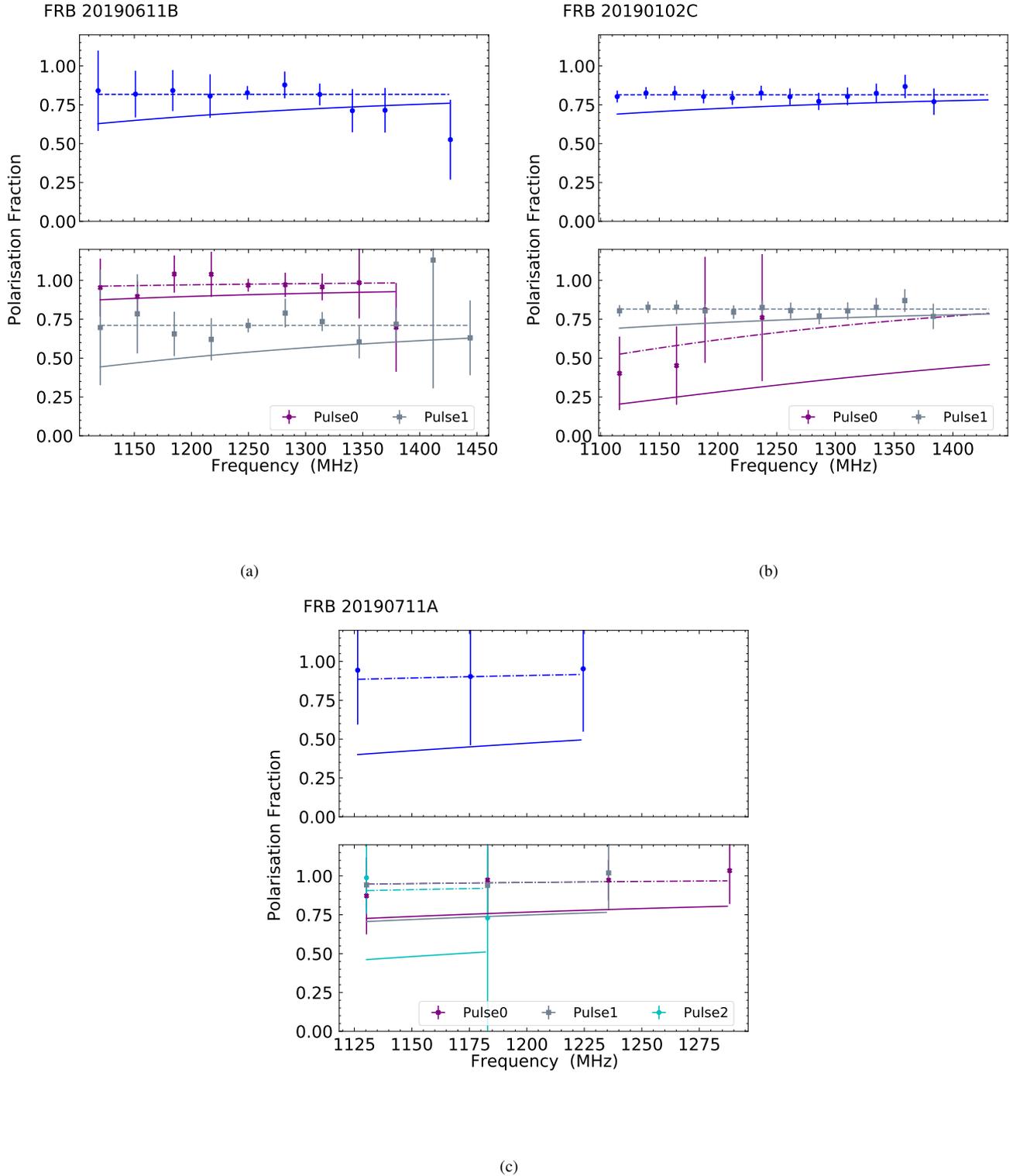

    
  \begin{subfigure}[t]{.49\linewidth}
    \centering\includegraphics[width=1\linewidth]{Figures/RM_scatter_fit_best_FRBs_2_panel_FRB190611.pdf}
    \caption{}
  \end{subfigure}
  \begin{subfigure}[t]{.49\linewidth}
    \centering\includegraphics[width=1\linewidth]{Figures/RM_scatter_fit_best_FRBs_2_panel_FRB190102.pdf}
    \caption{}
  \end{subfigure}\\
  \begin{subfigure}[t]{.49\linewidth}
    \centering\includegraphics[width=1\linewidth]{Figures/RM_scatter_fit_best_FRBs_FRB190711.pdf}
    \caption{}
  \end{subfigure}
\caption{Fractional linear polarisation of FRBs having multiple components (FRBs 20190611B, 20190102C, and 20190711A) across frequency. For each FRB, we show the model fit for the fractional linear polarisation, which has the highest evidence in dashed lines. The top panel shows the linear polarisation fraction for integrated pulse, and the bottom panel shows the linear polarisation for each sub-pulse. The preferred model in each case is shown in Table \ref{tab:evidence_parameters}. 
The solid lines denote the 95\% confidence upper limits derived from the modified Burn's law for sub-bursts and integrated bursts. Some bursts (e.g., Pulse 0 in 20190611B, Pulse 0 in 20190102C) have polarisation information in limited bandwidth due to lower polarised S/N - which are rejected due to de-biasing, hence polarisation information in the higher frequency band is missing.}
\label{fig:L_I_multi}
\end{figure*}

\begin{figure*}
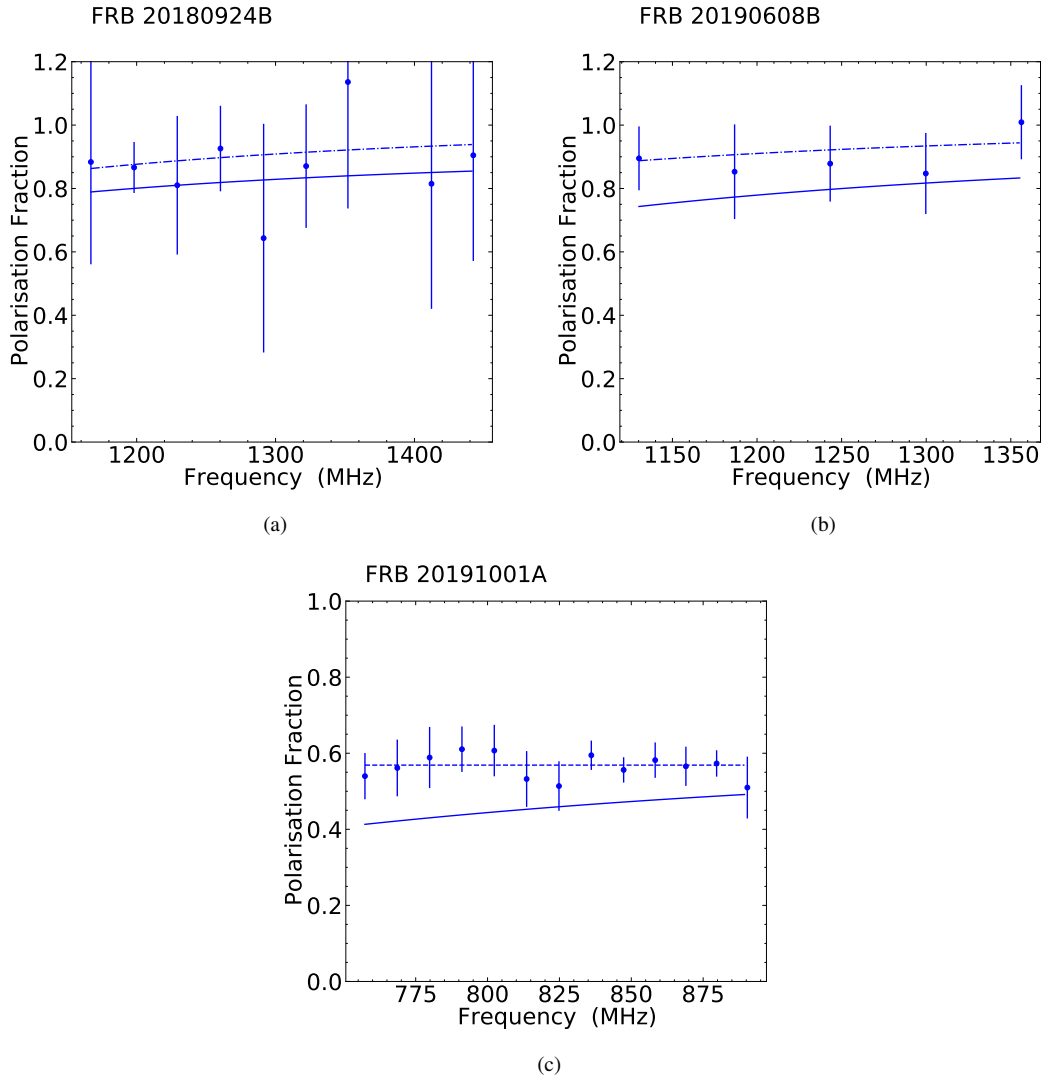

    \begin{subfigure}[t]{.4\linewidth}
        \centering\includegraphics[width=1\linewidth]{Figures/RM_scatter_fit_best_FRBs_FRB180924.pdf}
        \caption{}
    \end{subfigure}
    \begin{subfigure}[t]{.4\linewidth}
        \centering\includegraphics[width=1\linewidth]{Figures/RM_scatter_fit_best_FRBs_FRB190608.pdf}
        \caption{}
    \end{subfigure}\\
    \begin{subfigure}[t]{.4\linewidth}
        \centering\includegraphics[width=1\linewidth]{Figures/RM_scatter_fit_best_FRBs_FRB191001.pdf}
        \caption{}
    \end{subfigure}      

    \caption{Fractional linear polarisation of FRBs showing single components (FRBs 20180924B, 20190608B, and 20191001A) across frequency. The model fit for the fractional linear polarisation having the highest evidence is shown in dashed lines. Table \ref{tab:evidence_parameters} shows the preferred model in each case. The solid lines denote the 95\% confidence upper limits derived from the modified Burn's law for each FRB.}
    \label{fig:L_I_continued}
\end{figure*}


\renewcommand{\arraystretch}{1.5}
\setlength{\tabcolsep}{1pt}
\begin{table*}
\begin{tabular}{|c|c|r|r|r|r|r|r|}
			\hline
			TNS & 
			Sub-burst & 
			\thead{Burn's law \\ log$_{10}$ B} & 
			\thead{Const. Polarisation \\ log$_{10}$ B} & 
			\thead{Mod. Burn's law\\ log$_{10}$ B} & 
			\thead{\sigmaRM\\ \RMunits} & 
			\thead{$\sigma_{\rm RM}^\prime$\\ \RMunits} & \thead{P$_{int}$}\\
			\hline
			FRB 20180924B &  & 0.470 & \textit{0.88} & -0.16 & <5.17 & <4.01 & 0.91$^{+0.06}_{-0.07}$ \\
			FRB 20190102C &  & 8.140 & \textit{9.43} & 8.73 & <5.52 & <4.39 & 0.84$^{+0.05}_{-0.03}$ \\
			       & sub-burst 0 & \textit{-1.07} & -1.42 & -1.57 & <9.85 & <11.02 & 0.74$^{+0.17}_{-0.2}$ \\
			       & sub-burst 1 & 8.19 & \textit{9.42} & 8.74 & <5.5 & <4.33 & 0.84$^{+0.05}_{-0.03}$ \\
			FRB 20190608B &  & \textit{1.85} & 1.46 & 0.72 & <5.42 & <4.81 & 0.93$^{+0.05}_{-0.06}$ \\
			FRB 20190611B &  & 2.95 & \textit{3.26} & 2.60 & <6.41 & <5.48 & 0.85$^{+0.06}_{-0.04}$ \\
			       & sub-burst 0 & \textit{3.61} & 3.01 & 2.08 & <3.58 & <3.15 & 0.97$^{+0.02}_{-0.03}$ \\
			       & sub-burst 1 & 2.62 & \textit{2.82} & 2.41 & <8.27 & <7.28 & 0.77$^{+0.11}_{-0.06}$ \\
			FRB 20190711A &  & \textit{-0.16} & -0.46 & -1.04 & <8.64 & <8.93 & 0.85$^{+0.1}_{-0.2}$ \\
			       & sub-burst 0 & \textit{0.84} & 0.33 & -0.44 & <5.51 & <5.09 & 0.94$^{+0.04}_{-0.07}$ \\
			       & sub-burst 1 & \textit{0.53} & 0.14 & -0.63 & <5.7 & <5.18 & 0.92$^{+0.05}_{-0.09}$ \\
			       & sub-burst 2 & \textit{0.01} & -0.28 & -0.87 & <7.92 & <7.98 & 0.86$^{+0.1}_{-0.17}$ \\
			FRB 20191001A &  & 4.82 & \textit{8.86} & 7.89 & 4.1$^{+0.09}_{-0.10}$ & <2.8 & 0.6$^{+0.06}_{-0.03}$ \\
			\hline
\end{tabular}
\caption{\centering Model comparison for the ASKAP FRB sample.   Evidences are calculated for individual and integrated bursts. Higher values indicate preference for the model. For P$_{int}$ 68\% confidence intervals are reported, and for \sigmapRM\, and \sigmaRM, 95\% CI upper limits are reported, except for FRB 20191001A. The preferred model for each burst is shown in italic font.}
\label{tab:evidence_parameters}
\end{table*}


With the exception of FRB 20191001A, which slightly favours the constant polarisation model over Burn's depolarisation model, the bursts in our sample do not show strong evidence for or against any of the models.  The posterior probability distributions for all the models for FRB 20191001A are shown in Figure \ref{fig:191001_posterior}.
The preferred model fits for the linear polarisation fraction for each FRB sub-burst and integrated burst are shown in Figure \ref{fig:L_I_multi} and \ref{fig:L_I_continued}. 
While some of the components of some of the bursts show hints of depolarisation (the first sub-burst  of FRB 20190102C), the low S/N and fractional bandwidth results in inconclusive evidence.

\section{Discussion and Conclusions}
\label{sec:Discussion}

\begin{figure*}
    {\includegraphics[width=2\columnwidth]{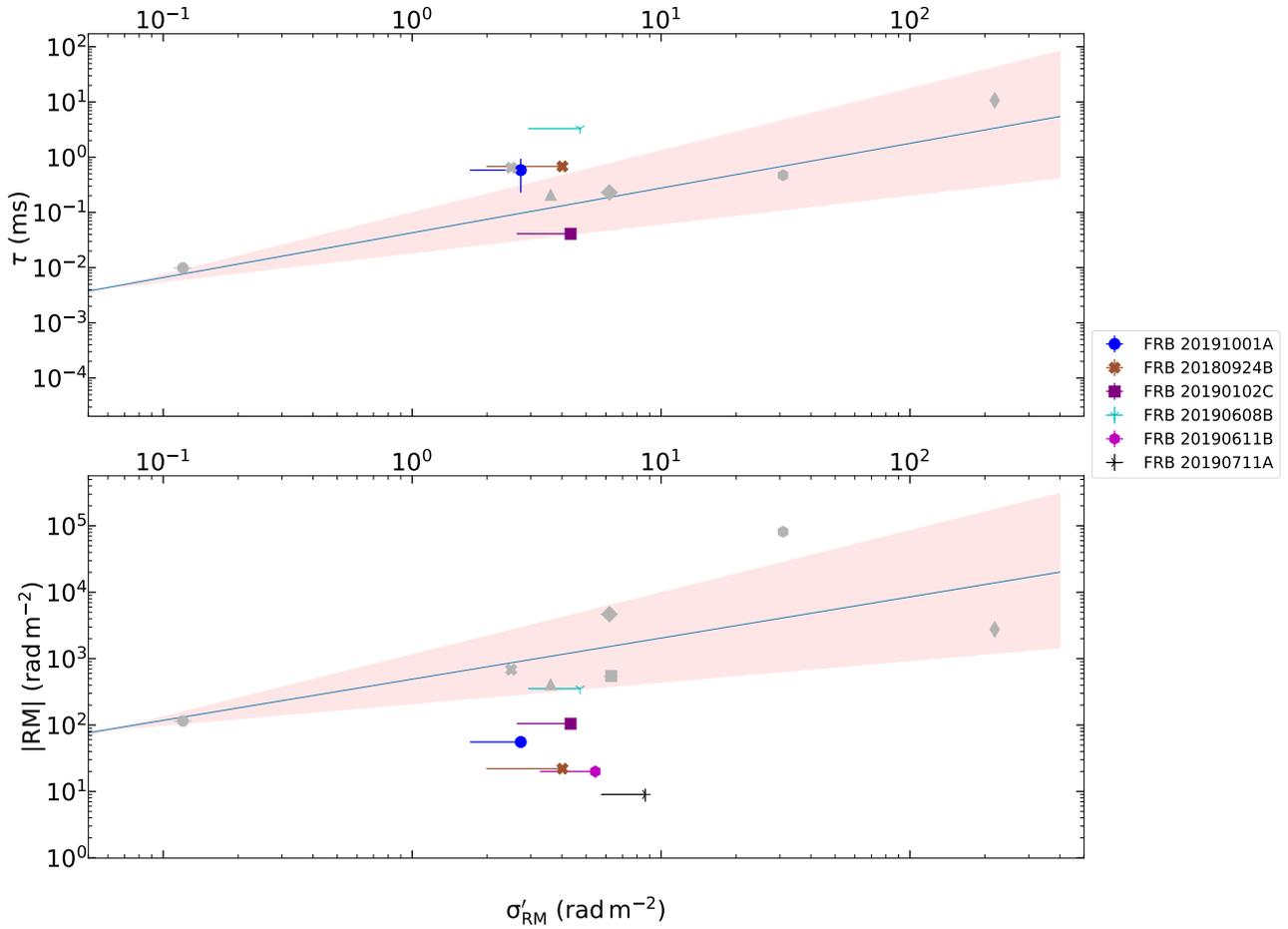}}
    \caption{Comparison of \sigmapRM, pulse broadening and RM. The conservative \sigmapRM\ 95\% confidence upper limits are derived from fitting the polarisation profiles of the individual bursts with modified Burn's law. The grey points are previously reported repeating FRBs. The scattering times for all the bursts have been scaled to 1271 MHz, assuming a $\nu^{-4}$ dependence of scattering timescale. The blue lines are the correlation fit previously reported in \protect\cite{Feng}. The red region indicates the 1-$\sigma$ uncertainty on the linear model fit from \protect\cite{Feng}. The uncertainties in RM are too small to see by eye.}
    \label{fig:correlations}
\end{figure*}

We consider five non-repeating and one repeating FRB detected by ASKAP and place limits on both \sigmaRM and \sigmapRM\ for each integrated burst and sub-burst (see Section \ref{sec:Depol_model} and Table \ref{tab:evidence_parameters}). We use \sigmapRM\ for our further discussion, as \sigmapRM\ conservatively accounts for potentially lower intrinsic linear polarisation fraction.

While we cannot determine conclusively whether or not the FRBs in our sample show spectral depolarisation, we can place useful constraints on the presence of Burn's depolarisation and compare our limits to the measurements of \sigmaRM\ published in the literature (shown in Figure \ref{fig:correlations}).
The upper limits on the derived \sigmapRM\ are consistent with \cite{Feng} \sigmaRM and RM  relationship, but not very constraining. 
The relationship between \sigmapRM and RM is consistent only if the values of \sigmapRM are $\sim$2 orders of magnitude smaller than our limits. 
If the model is applicable for non-repeating FRBs, then the depolarisation would only be observable by ASKAP for events with orders of magnitude higher RM or (if non-repeating FRBs inhabit low-RM environments) detected with other instruments at lower frequencies. 
The latter would suggest environments that are much less magnetoionically complex.



Our upper limits on \sigmapRM, however are inconsistent with the relationship between \sigmapRM and $\tau_s$ observed in repeating FRBs.
This indicates that even if a circumburst environment with similar properties to that inferred for repeating FRBs by \cite{Feng} exists, an additional source of scattering would be needed elsewhere along the line of sight.
In this scenario, the scatter broadening could be attributed to the host-galaxy ISM and not the circumburst environment. This is in contrast to the repeating FRB sources, for which \cite{Feng} concluded that the temporal scattering and \sigmapRM originated from the same inhomogeneous magnetoionic environment.
Alternatively, if a large fraction of scattering is being caused by the circumburst media, and not the host galaxy ISM, the inconsistency in the relationship between \sigmapRM and $\tau_s$ for repeating and non-repeating FRBs indicate a relatively less magnetised, but equally turbulent and dense, circumburst media when compared with the repeating FRBs from \cite{Feng}. 
The relationship between temporal scattering and depolarisation has been extensively studied in \cite{Yang_2022} for various physical scenarios in repeating FRB sources. In the case of shock-compressed magnetised plasma \cite{Yang_2022} predicts the \sigmaRM-$\tau_s$ relationship to be \sigmaRM$\propto$$\tau_{s}^{0.65-0.83}$ , which appears to be closer to the scaling index reported by \cite{Feng} for repeating sources. In general, the \sigmaRM-$\tau_s$ scaling depends on various physical parameters, including the radius of the assumed turbulence scale R and B$_{||}$, with lower B$_{||}$ leading to a shallower index of scaling or no correlation in \sigmaRM-$\tau_{s}$ which could be a characteristic of one-off FRBs. A similar correlation in the current set of one-off FRBs is not explored due to the lack of detection of depolarisation in any non-repeating FRBs. To robustly test for a different relationship between \sigmaRM and RM for one-off bursts, FRBs with comparable RMs to that of repeating FRBs are essential, or the detection of depolarisation (potentially using lower-frequency detection of one-off FRBs).  The largest magnitude RM reported in a one-off FRB is only 353 \RMunits\ for FRB 20190608B \citep{Day_2020}.

\begin{figure}
    \includegraphics[width=1\columnwidth]{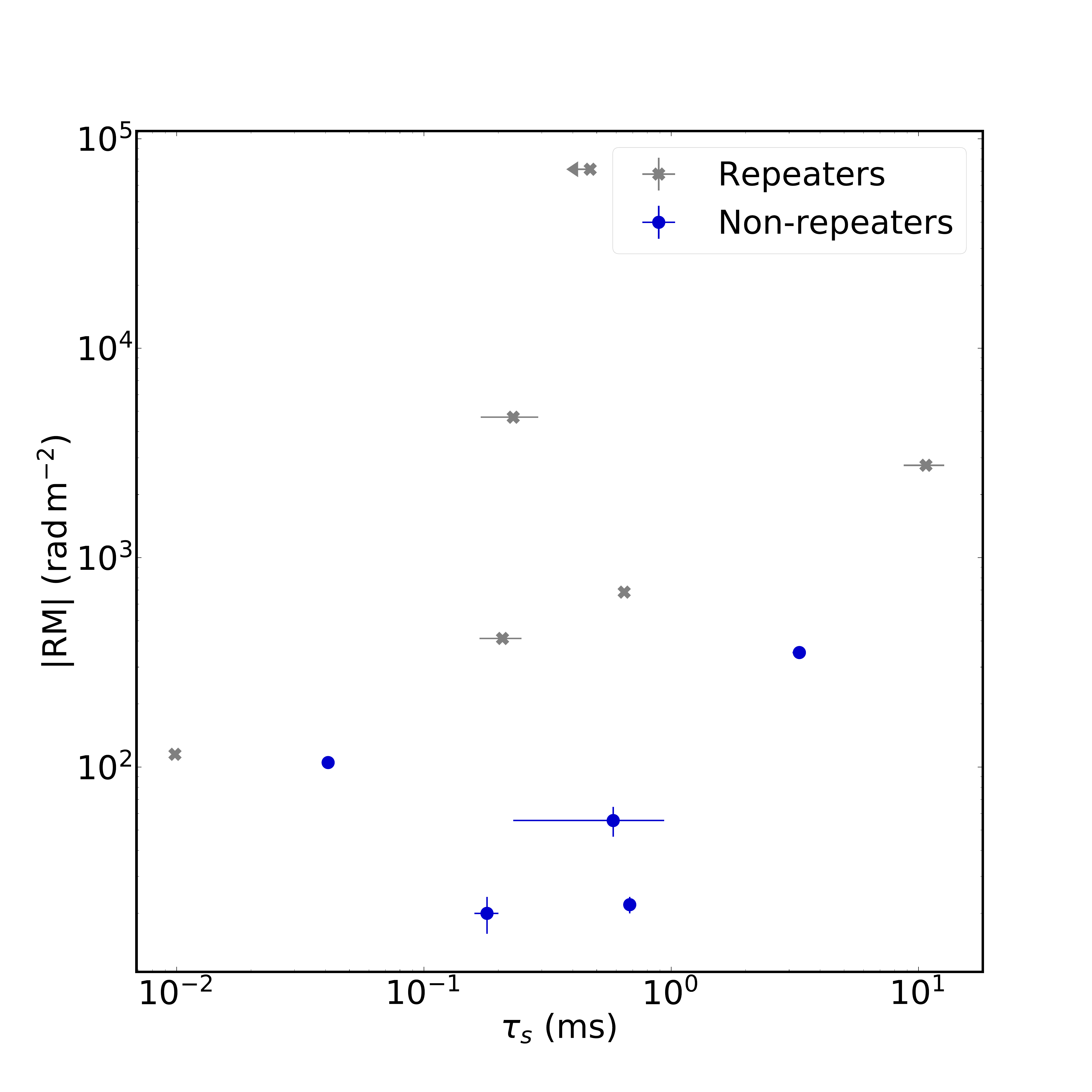}
    \caption{RM-$\tau_s$ relationship for repeaters reported in \protect\cite{Feng}, and our sample of non-repeaters. The $\tau_s$ for both repeaters and non-repeaters have been scaled to 1271 MHz, assuming a $\nu^{-4}$ dependence of the scattering timescale. The non-repeating FRB sample clearly shows a lower RM than that of the repeating FRBs for any given $\tau_s$.}
    \label{fig:RM_tau}
\end{figure}

The  properties of host galaxies of localised repeating and (apparently) non-repeating FRB sources do not show a strong distinction in global properties \citep{Bhandari_2022}. This provides potential evidence that we might expect similar lines of sight through bulk interstellar media, hence comparable ionised ISM for both the samples.
As such, we might expect similar $\tau_s$ for both repeating and non-repeating FRBs but the \sigmapRM\ to be independent and likely caused due to local environment.   
In this case, we would not expect there to be a positive correlation between $\tau_s$ and \sigmapRM\ for non-repeaters, as observed in repeating FRBs. Further, Figure \ref{fig:RM_tau}, which shows the relationship between RM and $\tau_s$, suggests similar $\tau_s$ but lower RMs for non-repeaters, which supports the scenario of less magnetised environments for non-repeaters. However, we caution the reader that a more extensive sample set of non-repeaters with polarisation measurements are required to confirm this.
     
While the differences in circumburst environments could be the result of non-repeating and repeating FRBs having different progenitors \citep[e.g.,][]{Caleb_2019, Cui_2022}, it is also possible that they have the same progenitor, with non-repeating FRBs sourced by an older population that has become less magnetoionically active in the later stages of its evolution.

Furthermore, for repeating FRBs such as FRBs 20201124A and 20190520B, depolarisation and RM variation have been speculated to be due to a magnetar main sequence Be-type star binary system, containing a decretion disk \citep{Dai_S, Wang_Y}, similar to the B1259-63 system \citep{Johnston}. However, in the case of non-repeating FRBs, since the follow-up observations are inherently unachievable, it necessitates future wideband polarimetric observations to correctly characterise the depolarisation from a single burst and rule out possible progenitor models based on depolarisation characteristics.    

\subsection{Effect of bandwidth in model selection}
The combination of the bandwidth of the instrument, observing frequency, and the S/N of the burst can impact our ability to constrain spectral depolarisation. While this is more pronounced in the case of repeating FRBs, which tend to be band-limited in comparison to that of one-off FRBs \cite[][]{Pleunis21}, by using follow-up observations at different frequencies, the polarisation properties can be measured over a range of frequencies, leading to a robust selection of a depolarisation model and an estimate of \sigmapRM. 
This is difficult for a one-off FRB with a limited instrumental bandwidth. In the case of the ASKAP non-repeating FRBs, a maximum of 336 MHz of the spectral window is achievable. However, even with this limited bandwidth, a better estimation of the parameters and the model is possible with bursts for which rapid exponential suppression of linear polarisation occurs within the band.
For ASKAP, assuming 1271 MHz and 824 MHz as the central frequencies and a 100\% intrinsic polarisation fraction, it will be most sensitive to a \sigmapRM\ of 12.68 \RMunits\ and 5.35 \RMunits\, respectively. 

The ability to measure $\sigma_{\rm RM}$ depends on the S/N ratio of the burst.
With a burst of S/N 61 (e.g., the highest S/N burst in our sample, reported by FRB 20191001A {\bf \citep{Bhandari}}), which corresponds to an uncertainty of 30\% in the linear polarisation fraction within a 1 MHz channel, over a bandwidth of 336 MHz and a center frequency of 824 MHz, no model discrimination would be possible above \sigmapRM of 8 \RMunits, and below 4 \RMunits.
Additionally, a broadband burst detected by the Ultra Wideband Low (UWL) receiver on Parkes \citep{Parkes_UWL} with an uncertainty of 10\% in linear polarisation, for an RM-corrected\footnote{rate in full Stokes data mode providing a typical spectral resolution of 0.5 MHz. We assume here that the we correct for RM before averaging the finer channels together. 
We choose 10\% uncertainty and  128 MHz as the channel width to realistically model a potential burst to have a total polarised SNR of 55.} 128 MHz sub-band in 3328 MHz of bandwidth, and a corresponding burst S/N of 55 should be able to discriminate depolarisation models across \sigmapRM of 3 \RMunits and 81 \RMunits.
Thus future wide-band polarimetric depolarisation studies, or multi-frequency simultaneous detection of bursts from different instruments, are required to better model the depolarisation behaviour, especially for non-repeaters.

\section*{Acknowledgements}

PAU and RMS acknowledges support through Australia Research Council Future Fellowship FT190100155.  RMS and ATD acknowledge support through Australian Research Council Future Fellowship FT220102305.  KG and ATD acknowledge support through DP200102243. 
SB is supported by a Dutch Research Council (NWO) Veni Fellowship (VI.Veni.212.058). This scientific work uses data obtained from Inyarrimanha Ilgari Bundara / the Murchison Radio-astronomy Observatory. We acknowledge the Wajarri Yamaji People as the Traditional Owners and native title holders of the Observatory site. The Australian SKA Pathfinder is part of the Australia Telescope National Facility (\href{https://ror.org/05qajvd42}{https://ror.org/05qajvd42}) which is managed by CSIRO. Operation of ASKAP is funded by the Australian Government with support from the National Collaborative Research Infrastructure Strategy. ASKAP uses the resources of the Pawsey Supercomputing Centre. Establishment of ASKAP, the Murchison Radio-astronomy Observatory and the Pawsey Supercomputing Centre are initiatives of the Australian Government, with support from the Government of Western Australia and the Science and Industry Endowment Fund.

This work makes use of \code{BILBY} \citep{Ashton}, \code{MATPLOTLIB} \citep{Hunter:2007} and \code{NUMPY} \citep{harris2020array} software packages.

\section*{Data Availability}
No new data were generated in this work.



\bibliographystyle{mnras}
\bibliography{example}



\appendix

\section{Burn's depolarisation Model}
\label{apx:foreground_depolarisation}
The Faraday Dispersion Function described in \cite{Burns_1966} relates the complex linear polarisation function P($\lambda^2$) to a Faraday dispersion function (FDF, F$(\phi)$) using a Fourier transform. It describes the polarisation behaviour of the source across wavelength due to the Faraday dispersion caused by the intervening media. It can explain the decrease in observed polarisation fraction in wavelength for sources which are expected to have a polarisation fraction independent of frequency \citep{Le_Roux}. The FDF relates the complex polarised brightness of the source at unit Faraday depth $\phi$:
\begin{equation}
    P(\lambda^2) = \int_{-\infty}^{\infty} F(\phi) \, \exp{\left ( 2i\phi\lambda^2 \right )} d\phi,
    \label{eq:FDF}
\end{equation}
where $P(\lambda^2) = Q(\lambda^2)+i U(\lambda^2)$ is the complex polarisation function and $\lambda$ is the wavelength.
$\phi(r)$ is the Faraday depth of the emission: 
\begin{equation}
    \phi(r) = 0.81 \int_{r}^{0} n_e {\bf B} \cdot d  {\bf r},
    \label{eq:Faraday_depth}
\end{equation}
where $n_e$ is the electron density in cm$^{-3}$, $B$ is the magnetic field in ${\rm \mu}$G and $dr$ is the infinitesimal path length in parsec. In this case, the units of  $\phi(r)$ are \RMunits. 
Depolarisation can also arise due to beam depolarisation. However, this is not applicable to FRBs which are point sources relative to the beam of the telescope. Using Equation (\ref{eq:FDF}), the FDF can be calculated assuming $P(\lambda^2)$ to be Hermitian, which was successfully demonstrated for early observations of Crab nebula \citep{Gardner, Burns_1966}.
There are two possible sources for depolarisation, internal and external to the emission region. 
If the sign or strength of the parallel component of the magnetic field changes across the source emission region, there will be depolarisation, as the emission will arise from different rotation measures. This is less likely to be the case for FRBs, as the emission is constrained to come from a region smaller than 300$\,$km\,$\Gamma w_{i,-3}$, where $w_{i,-3}$ is the intrinsic burst width measured in ms, and $\Gamma$ is the (currently theoretically unconstrained) Lorentz factor of the source plasma.  The emitting region is also likely to reside in a relativistic plasma, which does result in conventional Faraday rotation. 

The second component of the depolarisation relates to the foreground turbulent RM scatter, which is separate from the internal depolarisation. Considering the distribution of RM as a Gaussian random variable, the expectation of Equation \ref{eq:FDF} results in Equation \ref{eq:Burns_law}. 

A final contribution to depolarisation could be due to the polarised foreground emission. In our analysis, the polarised emission from a constant foreground is removed during the off-pulse subtraction of the Stokes spectrum.  

\bsp	
\label{lastpage}
\end{document}